\documentclass[12pt]{article}
\usepackage{psfig,float}
\begin{document}

\begin{center}
{\large{\bf THEORETICAL STUDY OF THE $\gamma \gamma \rightarrow $ MESON-MESON
REACTION}}
\end{center}

\vspace{1cm}

\begin{center}
{\large{J.A. Oller and E. Oset}}
\end{center}

\begin{center}
{\small{\it Departamento de F\'{\i}sica Te\'orica and IFIC\\
Centro Mixto Universidad de Valencia-CSIC\\
46100 Burjassot (Valencia), Spain}}
\end{center}

\vspace{3cm}

\begin{abstract}
{\small{
We present a unified theoretical picture which studies simultaneously
the $\gamma \gamma \rightarrow \pi^+ \pi^-$, $\pi^0 \pi^0$, $K^+ K^-$, 
$K^0 \bar{K}^0$, $\pi^0 \eta$ reactions up to about $\sqrt{s} = 1.4 \,\,
 GeV$ reproducing the experimental cross sections . The
present work implements in an accurate way the final state interaction
of the meson-meson system, which is shown to be 
essential in order to reproduce the data, particularly the $L = 0$ channel.
This latter channel 
is treated here following a recent theoretical work in
which the meson-meson
 amplitudes are well reproduced and the $f_0, a_0, \sigma$
resonances show up clearly as poles of the $t$ matrix. The present work,
as done in earlier ones,
also incorporates elements of chiral symmetry and exchange of vector and
axial resonances in the crossed channels, as well as a direct coupling to the
$f_2 (1270)$ and $a_2 (1370)$ resonances. We also evaluate the decay 
width of the $f_0 (980)$ and $a_0 (980)$ resonances into the $\gamma 
\gamma$ channel.}}  
\end{abstract}

\newpage

\section{Introduction}

The $\gamma \gamma \rightarrow$ meson-meson reaction [1--9]
provides interesting information concerning the structure of hadrons,
their spectroscopy and the meson-meson interaction, given the sensitivity 
of the reaction to the hadronic final state interaction (FSI)
\cite{9,10}. The main aim of the present work  is to offer a unified
description of  the different channels $\gamma \gamma \rightarrow M \bar{M}$,
where $M$ are the mesons of the lightest pseudoscalar octet
$(\pi, K, \eta$), concretely $\gamma \gamma \rightarrow \pi^0 \pi^0,
\pi^+ \pi^-, K^+ K^-, K^0 \bar{K}^0$ and $\pi^0 \eta$.
 We shall see, indeed, that the consideration of the
meson-meson interaction is essential in order to interpret the data up
to about $\sqrt{s} = 1.4$ $GeV$.

For the meson-meson interaction we are forced to take a theoretical 
framework which works up to these energies. The chiral
perturbation approach \cite{11,12,13,14} does not serve this purpose
since its validity is restricted to much lower energies. However, chiral
symmetry is one of the important ingredients when dealing with the
meson-meson interaction, and its potential to predict and relate 
different processes is not restricted to the perturbative regime, as shown 
in \cite{15,Truong,Zahed}. In ref. \cite{15} we developed a non-perturbative
theoretical scheme which takes chiral symmetry into consideration and
allows one to obtain the meson-meson amplitudes quite accurately up
to about $\sqrt{s} = 1.2$ $GeV$. The scheme starts from the lowest
order chiral amplitudes which are used as potentials in coupled channel
Lippmann Schwinger equations with relativistic meson propagators. Only
one parameter is needed in this approach, a cut off in the loops, of the order
of 1 $\, GeV$, as expected from former calculations of the scale of the
chiral symmetry breaking $\Lambda_\chi$ \cite{16},
which plays a similar role as a scale of energies as our cut off. The
scheme implements unitarity in coupled channels and
produces the $f_0  (980), a_0 (980)$ and $\sigma$ resonances with
their corresponding masses, widths and   branching ratios. In ref. \cite{15}
the $L = 0$ and $T = 0$, $T = 1$ channels were studied. Here we extend the
model to account for the $L = 0$, $T = 2$ channel which is also needed
in the present problem.

In addition to the $f_0 (980), a_0 (980)$ and $ \sigma$ resonances, which
come up naturally in the approach of ref. \cite{15}, we introduce 
phenomenologically the $f_2 (1270) $ and $a_2 (1320)$ resonances in
the $L = 2$ channel in order to account for the upper part of the
energy spectrum.

Another relevant aspect of this reaction, which has been reported previously,
is the role of the vector and axial resonance exchange in the $t$, $u$
channels \cite{17}. We shall also take this aspect into consideration
using the 
effective vertices of refs. \cite{17,18}. The $f_0$ and $a_0$ resonances 
appear  also as
singularities  of the $\gamma \gamma \rightarrow $ meson-meson amplitude,
through the meson-meson amplitudes \cite{15}.
We also evaluate the partial decay widths of the $f_0$ and $a_0$
resonances into the $\gamma \gamma$ channel.

Given the relevance of the meson-meson interaction in the $\gamma \gamma
\rightarrow M \bar{M}$ process, and the accessibility of the low
energy regime in $\gamma \gamma \rightarrow \pi \pi $, this reaction has 
been a testing ground of the techniques
of chiral perturbation theory ($\chi P T$), particularly in the
$\gamma \gamma \rightarrow \pi^0 \pi^0$ case where there is no direct
coupling of the photons to the $\pi^0$ and the first contribution involves
one loop with no counterterms at order $O (p^4)$ \cite{20,21}.

The disagreement of the $\chi P T$ results for 
$\gamma \gamma \rightarrow \pi^0 \pi^0$ at one loop with the Crystal
Ball data \cite{1} motivated calculations up to order $O (p^6)$ \cite{22}
where the agreement with this experiment was improved.

Evaluations in $\chi P T$ for the $\gamma \gamma \rightarrow \pi^+
\pi^-$ case have also been performed to order $O (p^4)$ \cite{20} and
to order $O (p^6)$ \cite{23}. 

Improvements beyond the $O (p^4)$ results using dispersion
relations and resonance exchange, and matching the results to those
of $\chi P T$ at low energies, have also been performed in
ref. \cite{18}.

A different approach is developed in \cite{Zahed,Zahed2}, where a 
master formula for chiral symmetry breaking is deduced for the $SU(2)$ case 
which allows to relate the $\gamma \gamma \rightarrow \pi\pi$ reaction with other 
physical processes in a non perturbative way. In order to obtain 
numerical results the form factors and correlation functions appearing 
in the formalism must be modelled, and this is done making use essentially 
of the resonance saturation hypothesis.

Another step forward is given in ref. \cite{25}, where calculations
with two loops including counterterms to all orders in the leading 
$1/N_c$ contribution are performed within the extended 
Nambu-Jona-Lasinio model.

A different analysis, more phenomenological, of these amplitudes
is also done in \cite{26,27,28} by imposing basic symmetries as 
unitarity, analyticity and using experimental phase shifts, resonance
parametrization, etc.

There is also a controversial point related to the meson-meson and $\gamma
\gamma \rightarrow$ meson-meson amplitudes, which is the hypothetical
existence of a broad scalar resonance in the $T = 0$ channel 
denoted $f_0 (1100)$ in ref. \cite{27}. This resonance 
 does not show up in $\gamma \gamma \rightarrow
\pi^0 \pi^0$, as it is stressed in ref. \cite{10}.
 In our case, both in the former work \cite{15} on the 
meson-meson interaction and in the present one on the $\gamma \gamma
\rightarrow M \bar{M}$ reaction, the experimental results are
well reproduced without the need of introducing this resonance, which
unlike the $f_0$ and $\sigma$, which appear naturally in the theoretical
framework, does no show up in the meson-meson amplitude of ref.
\cite{15}.

In the case of the $\gamma \gamma \rightarrow \pi^0 \eta$ we 
reproduce the experimental results of refs. \cite{oest,8}
 in terms of our S-wave calculated amplitude, which includes the peak of the 
$a_0(980)$, plus the $a_2(1320)$ resonance contribution without the need 
to include an extra background from a hypothetical $a_0(1100-1300)$ 
resonance, see ref. \cite{10}, which was also not needed in ref. \cite{15}.

A novel result of the present work is the reproduction of the
$\gamma \gamma \rightarrow K^+ K^-$ cross section. This reaction
was particularly problematic since the Born term  largely
overestimates the experimental data from threshold on. The need of 
a theoretical mechanism to drastically reduce this background in the
reaction has been advocated \cite{10}, without a solution, so far, to
this problem.
As we shall see later on, this reduction is automatically obtained
in our work
in terms of the final state $M \bar{M}$ interaction.

As with respect to the $\gamma \gamma \rightarrow K^0 \bar{K}^0$
we obtain a small background, as expected \cite{10}, but it is not a 
consequence of the lack of the Born term but rather a result of
cancellations
between terms of the order of magnitude of the Born term
in the $\gamma \gamma \rightarrow K^+ K^-$ amplitude.

\section{Vertices in the $\gamma \gamma \rightarrow M \bar{M}$
reaction}

We proceed now to evaluate the $\gamma \gamma \rightarrow M \bar{M}$
amplitudes to tree level. The final amplitudes will be constructed from these
ones including the final state interaction of the $M \bar{M}$ system.

\begin{figure}[h]
\centerline{
\protect
\hbox{
\psfig{file=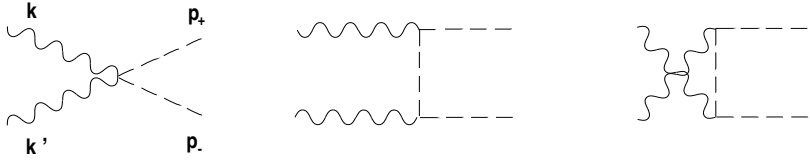,width=0.7\textwidth,angle=-90}}}
\caption{Born term amplitude for $\gamma\gamma \rightarrow M^+M^-$. $k$ and 
$k'$ are the momenta of the incoming photons and $p_+$($p_-$) the momentum 
of the positively(negatively) charged meson.}
\end{figure}

For the charged mesons $\pi^+ \pi^-$ and $K^+ K^-$ we have the Born term

\begin{equation}
t_B = - 2 e^2 \{ \epsilon_1 \cdot \epsilon_2 - 
\frac{p_+ \cdot \epsilon_1 \; p_- \cdot \epsilon_2}{p_+ \cdot k} -
\frac{p_+ \cdot \epsilon_2 \; p_- \cdot \epsilon_1}{p_+ \cdot k'} \}
\end{equation}

\noindent
following the notation of fig. 1 for the momenta of the particles. We 
use the gauge
 $\epsilon_1 \cdot k = \epsilon_2 
\cdot k = \epsilon_1 \cdot k' = \epsilon_2 \cdot k' = 0$,  where $\epsilon_1,
\epsilon_2$ are the photon polarization vectors.

Following the work of refs. \cite{17,18,30} we consider the exchange
of the octets of vector and axial resonances in the $t$
and $u$ channels. For real photons the
vector sector is dominated by the $\omega$
 exchange, since the $\omega$ coupling to $\gamma \pi^0$ 
 is about one order of magnitude bigger than the one of the $\rho$ and
$K^*$. We include $\omega$ and $\rho$ exchange in our calculations,
although the role of the $\rho$ is negligible, in order to compare our
results with those of ref. \cite{18} (see fig. 2). In this way we take 
into account the contribution of the left hand cut, to which the Born 
term also contributes \cite{18}. Thus we are considering the relevant 
elements of crossing symmetry which are important to relate the present 
process to Compton scattering on mesons and their related polarizabilities 
\cite{18}.

\begin{figure}[h]

\centerline{
\protect
\hbox{
\psfig{file=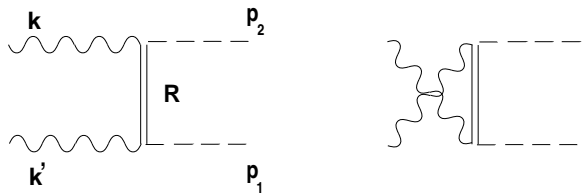,width=0.5\textwidth,angle=-90}}}
\caption{Tree level amplitude for  $\gamma\gamma\rightarrow M_1M_2 $ through 
the exchange of a resonance $R$(axial or vectorial) in the t,u channels. 
$k$ and $k'$ are the momenta of the incoming photons and $p_1$, $p_2$ are 
the momenta of the final mesons.}
\end{figure}

The contribution from the axial resonance exchange is given by 
\cite{18}

\begin{equation}
t_A ' = 4 \pi \alpha 
\frac{F_A^2}{f^2} \,  \epsilon_1 \cdot \epsilon_2 \, \,
k \cdot k' [ \frac{1 + \frac{p_1 (k - p_1)}{m_A^2}}{(k - p_1)^2-m_A^2} 
+ \frac{1 + \frac{p_1 (k' - p_1)}{m_A^2}}{(k' - p_1)^2-m_A^2}]
\end{equation}

\noindent
where $f$ is the pion decay constant, $f = 93 \, MeV$. The coupling
$F_A$ can be related to the phenomenological $L^r_9 +
L^r_{10}$ coefficients of the order $O (p^4)$ chiral Lagrangian \cite{11},
by using the KSFR relation \cite{31}, $m_A = \sqrt{2}
m_\rho$, and one obtains \cite{11}

$$
\frac{F_A^2}{4 m_A^2} = L_9^r + L_{10}^r = (1.4 \pm 0.3) \cdot 10^{-3}
$$

In the following we shall take the central value for $L_9^r+L_{10}^r$.

The $\rho \pi \gamma$ and $\omega \pi \gamma$ vector couplings are of the
form

\begin{equation}
t (V \rightarrow \gamma M) = i 2 e \sqrt{R_V} \epsilon^{\mu \nu \alpha
\beta} \partial_\mu V_\nu \epsilon_\alpha p_{M, \beta}
\end{equation}

\noindent
with $R_V$ fitted to the  partial decay width $V \rightarrow
\gamma M$.
We obtain $R_\omega = 1.35 \,\;  GeV ^{-2}$, $R_\rho = 0.12 \,\; GeV^{-2}$, 
which shows clearly the negligible role that the $\rho$ plays here (note 
that these coupling constants appear squared in the 
$\gamma \gamma \rightarrow M \bar{M}$ amplitudes as shown in fig. 2). 
From the coupling of eq. 
(3) one can easily evaluate the
$t_V$ amplitude and explicit expressions for it can be found in ref.
\cite{18} for on shell photons and mesons. The coupling of the $K^*
\rightarrow \gamma K$ is of the same order of magnitude as the one of
the $\rho$, which justifies neglecting it here.

We shall also need the off shell extrapolation and hence we give below 
the expression of the only amplitude that contributes in S-wave, which 
is the $ t^{'++}_R $ in helicity basis (see eq. 23)

\begin{equation}
t_R ^{'++}=i\frac{4 e^2 R_V}{(k-p_1)^2-m_V ^2} \left\{ \frac{1}{2}
(k \cdot 
k') |\vec{p}| ^2 _1 \sin^2 \theta - (k \cdot k')(p_1 p_2) + (k' p_1)(
k p_2) \right\} + \left\{ k \leftrightarrow  k' \right\}
\end{equation}

\hspace{-0.7cm}
where $ \theta $ is the angle between the photon of momentum $\vec{k}$ and the 
meson of momentum $\vec{ p_1} $ .

\section{Final state interaction corrections}

We separate contributions from the S and D-waves of the
$\gamma \gamma \rightarrow M \bar{M}$ amplitude.

\subsection{$S$-wave.}

The one loop contribution generated from the Born terms with 
intermediate charged mesons can be directly taken from $\chi P T$ 
calculations of the $\gamma \gamma \rightarrow \pi^0 \pi^0$ amplitude.

From refs. \cite{20,21}, taking for the moment only intermediate
charged pions, we have above the $\pi \pi$
threshold

\begin{equation}
t = \frac{\epsilon_1 \epsilon_2}{16 \pi^2} (\frac{2 e^2}{f^2}) (s - m_\pi^2)
\left\{ 1 + \frac{m_\pi^2}{s} \left[ ln
\left( \frac{1 + (1 - 4 m_\pi^2/s)^{1/2}}{ 1 - 
(1 - 4m_\pi^2/s)^{1/2}} \right) - i \pi \right]^2 \right\}
\end{equation}

\noindent 
with its corresponding analytical extrapolation below pion threshold.

By taking into account the fact that the $\pi^+ \pi^-
\rightarrow \pi^0 \pi^0$ amplitude is given in $\chi P T$ at
order $O (p^2)$ by

\begin{equation}
t_{\pi^+ \pi^- , \pi^0 \pi^0} = - \frac{s - m_\pi^2}{f^2}
\end{equation}

\noindent
we see that this meson-meson amplitude factorizes in eq. (5) with its
on shell value of eq. (6). Our contribution beyond this point is
to include all meson loops generated by
the coupled channel Lippmann Schwinger equations of ref. \cite{15}, in
which we also saw that the on shell meson-meson amplitude factorizes
outside the loop integrals. Schematically, the series of terms
generated is depicted in fig. 3. Hence, the immediate consequence of
introducing these loops is to substitute the on shell $\pi \pi$ amplitude
of order $O (p^2)$ in eq. (6) by our on shell meson-meson amplitude
evaluated in ref. \cite{15}. This result, which is an exact consequence
of the use of the approach of ref. \cite{15}, was suggested already in
ref. \cite{21} as a means to improve the results of $\chi P T$. The same 
conclusion about the factorization of the strong amplitude was reached in 
ref. \cite{Dobado} for the $SU(2)$ case and the large N limit ( N is the 
number of Nambu-Goldstone bosons).

\begin{figure}[h]

\centerline{
\protect
\hbox{
\psfig{file=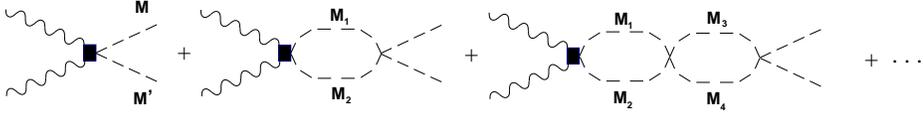,width=0.8\textwidth,angle=-90}}}
\caption{Diagrammatic series which gives rise to the FSI from a general 
 $\gamma\gamma\rightarrow M\,M'$ vertex.}
\end{figure}

The one loop contribution involving charged kaons
can be obtained from eq. (5) by changing the amplitude
$-(s - m_\pi^2)/f^2$ by the corresponding $K^+ K^- \rightarrow
\pi^0 \pi^0$ one and $m_\pi^2 \rightarrow m_K^2$ in the rest of the formula.
This is generalized to any meson-meson in the final state by changing the
corresponding meson-meson amplitude.

\begin{figure}[h]
\centerline{
\protect
\hbox{
\psfig{file=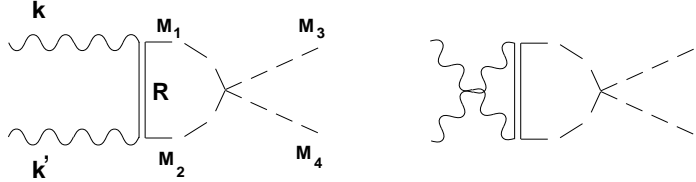,width=0.6\textwidth,angle=-90}}}
\caption{One loop correction for the  $\gamma\gamma\rightarrow M_3M_4$ tree 
level amplitude through the exchange of a resonance $R$ in the crossed 
channels. $M_1$ and $M_2$ are the intermediate states in the loop and 
$k$, $k'$ are the momenta of the incoming photons.}
\end{figure}

We can also apply a similar procedure to account for FSI in the terms
generated by vector and axial resonance exchange discussed in {\bf section 2}. 
For this purpose we take the diagrams of fig. 4 and then the contribution of
these  one loop corrections, plus the extra iterations in the meson-meson 
amplitude discussed above, can be easily taken
into account by means of

$$
t_{R, M_3 M_4} = \sum_{M_1 M_2} \tilde{t}_{R, M_1 M_2} t_{M_1 M_2, M_3 M_4}
$$

\noindent

\hspace{-0.7cm}where 

\begin{equation}
\tilde{t}_{R, M_1 M_2} =i \int \frac{d^4 q}{(2 \pi)^4} t'_R,_{M_1 M_2}
 \, \frac{1}{q^2 - m_1^2 + i \epsilon}
 \, \, \frac{1}{(P - q)^2 - m_2^2 + i \epsilon}
\end{equation}

\noindent
where once again the on shell strong meson-meson amplitude factorizes
outside the integral.

The electromagnetic amplitude has a different structure to the strong
one and must be kept inside the integral. In eq. (7) $P = (\sqrt{s},
 \vec{0})$ is
the total fourmomentum of the $\gamma \gamma$ system and $m_1, m_2$
the masses of the intermediate $M_1, M_2$ mesons. In addition, and in 
analogy to the work of ref. \cite{15}, the integral over $|\vec{q}|$ 
in eq. (7) is 
cut at $q_{max} = 0.9 \, GeV$.

One can justify the accuracy of factorizing the 
strong amplitude for the loops with crossed exchange of resonances.
 Take for instance, eq.(2) for the exchange of the axial 
resonance and assume the limit $M_A \rightarrow \infty$ . From this equation 
we see that in this limit $t'_A$ can be factorized outside the integral 
of eq. (7). Then if one takes the off shell part of the strong amplitude 
one cancells the propagators in eq. (7), after the $q^0$ integration, 
and obtains a result of the type 
$\Lambda^2 t'_A$ ($\Lambda$ is the cut off, $q_{max}$), which would 
renormalize the effective $t'_A$ amplitude and 
since we are taking the physical value for the coupling constants 
this term should be omitted. These arguments are shown 
in detail in ref. \cite{15}. A similar argumentation can be done for the 
exchange of vector mesons. Since we are dealing with real photons the 
intermediate axial or vector mesons are always off shell and the large 
mass limit is a sensible approximation. This implies that the error in 
the real part of the loops shown in fig. 4, in the way we estimate it, 
has an expansion in powers of $M_R^{-2}$ such that for $M_R \rightarrow 
\infty$ is zero. In this way if we call $\Delta _{\pi}$ the relative desviation 
between the exact value for the real part coming from eq. (5), $M_R=m_{\pi}$, 
and the value we would obtain following the procedure in eq. (7), then 
the relative error for a resonance mass $M_R$ 
would be $\Delta_R \simeq \Delta_{\pi} \bigl[ \frac{m_{\pi}}{M_R}
\bigr] ^2$ which results in uncertainties below the level of 5$\%$ for 
$M_R$ about 800 $MeV$.

 One of the limitations of the unitary method for 
meson-meson interaction of \cite{15} is the lack of crossing symmetry. The 
unitarization is done in the s-channel but not in the t or u channels. 
In practice this limitation means that one should not use crossing 
symmetry to relate crossed channels. Instead, when the amplitude 
of these crossed channels has to be evaluated one redefines the new 
s-channel and applies then the method. The left hand cut neglected in 
our procedure is expected to be less important at high energies 
because the physical energy is far away from the cut. On the other hand, at 
low energies the loops and counterterms are in any case less important 
and are dominated by those of the s-channel unitarity considered here. 
These arguments are qualitative but they have been put in quantitative 
form in \cite{Penni,Hannah} and the corrections are at the level of 
$2 \%$, even at energies near the two pion threshold. 

\subsection{S-wave $T = 2$ strong $\pi \pi$ amplitude}
In the work of ref. \cite{15} the $L = 0$ and $T = 0$,
$T = 1$ meson-meson amplitudes are evaluated. For the $\gamma \gamma
\rightarrow \pi \pi$ we need also the $T = 2$ channel. For this 
purpose we extend the work of ref. \cite{15} to the $T = 2$
channel. In this latter case we have only pions since 
$K \bar{K}$ or $ \pi^0 \eta $ does not couple to $T = 2$. We get then

\begin{equation}
t^{(T = 2)} = \frac{V^{(T = 2)}}{1 - G V^{(T = 2)}}
\end{equation}

\noindent
where from ref. \cite{15}

\begin{equation}
G (s) = \int_{0}^{q_{max}} dq \frac{q^2}{2 \pi^2} \, \frac{1}{\omega
(P^{02} - 4 \omega^2 + i \epsilon)}
\end{equation}

\noindent
with $\omega = (\vec{q}\,^2 + m^2_\pi)^{1/2}$. The potential $V^{(T = 2)}$
is given by

\begin{equation}
V^{(T =  2)} = \frac{1}{2} \, \frac{s - 2 m_\pi^2}{f^2}
\end{equation}

\noindent
where following again the notation of ref. \cite{15} we used the
``unitary'' normalization for the states

\begin{equation}
| \pi \pi, T = 2, T_3 = 0 > = - \frac{1}{\sqrt{12}} \, | \pi^+ (\vec{q})
\pi^- (- \vec{q}) + 
\pi^- (\vec{q}) \pi^+ (- \vec{q}) - 2 \pi^0 (\vec{q}) \pi^0
(- \vec{q}) >
\end{equation}

This normalization (with an extra normalization $(1/\sqrt{2}$)
is introduced in order to use the standard formula for the phase shifts
when using identical particles

\begin{equation}
t^{(T = 2)} = - \frac{8  \pi \sqrt{s}}{2i}  \, \frac{e^{2 i \delta} - 1}{p}
\end{equation}

\noindent
where $p$ is the CM momentum of the pion. The physical amplitude is given
by $t (\theta) + t (\pi - \theta)$ and in S-wave this amounts to multiplying
by a factor two the amplitude obtained in eq. (8).

In fig. 5 we show the phase shifts of $\pi \pi$ in $L = 0$ and $T = 2$ and
compare them with the experimental results of ref. \cite{32,33,34}. We see 
an agreement with the experimental results up to about $\sqrt{s} =0.8 
\,\, GeV$.  

\begin{figure}[h]
\centerline{
\protect
\hbox{
\psfig{file=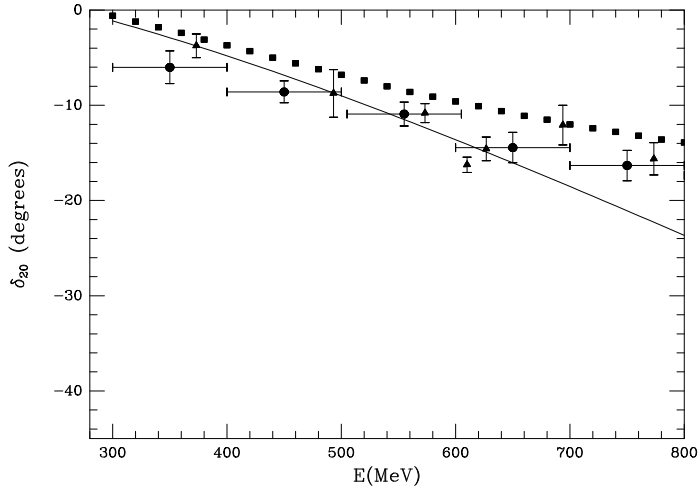,width=0.6\textwidth,angle=-90}}}
\caption{$\delta_{TL}=\delta_{20}$ phase shifts for the elastic $\pi\pi$ 
amplitude whith $L=0$ and $T=2$ from threshold to $0.8$ $GeV$. The squares 
are from \cite{32}, circles from \cite{33}  and the triangles from \cite{34}.}
\end{figure}

In order to have some more accurate results at energies higher than $\sqrt{s}
= 0.8 \,\, GeV$  we take the experimental phase shifts. This has irrelevant
consequences in the $\gamma \gamma \rightarrow \pi^+ \pi^-$ amplitude
and introduces changes of the order of 10$\%$ at high energies in the
$\gamma \gamma \rightarrow \pi^0 \pi^0$ channel with respect to using
the $T = 2$ amplitude of our theoretical approach.

\subsection{D-wave contribution}

As one can see from ref. \cite{26} (see fig. 7 of this reference) the
Born term in $\gamma \gamma \rightarrow \pi^+ \pi^-$ is given 
essentially by the $(0, 0)$ and $(2, 2)$ partial waves, in
$(J, \lambda)$ notation, where $J$ is the angular momentum and
$\lambda$ is the difference of helicities of the two photons.

The $(0,0)$ component has already been taken care of in {\bf section 3.1}. For
the $(2, 2)$ component we take the results of ref. \cite{26}, obtained
using dispersion relations, and which are parametrized in the form

\begin{equation}
t_{BC}^{(2,2)} = \biggl[ \frac{2}{3} \, \chi_{22}^{T=0} \, e^{i \delta_{20}} +
 \frac{1}{3} \, \chi_{22}^{T=2} \, e^{i \delta_{22}} \biggr]\, t_B^{(2, 2)}
\end{equation}

\noindent
where $t_B^{(2, 2)}$ is the $(2, 2)$ component of the Born term, 
$\delta_{20}$ and $\delta_{22}$
are the phase shifts of $\pi \pi \rightarrow \pi\pi$ in $L = 2$, $T = 0$ 
and $T=2$ respectively. 
$\chi_{22}^{T=0}$ is a function which is approximately given by

\begin{equation}
\chi_{22} = 1 - (\frac{s}{s_0})^2
\end{equation}

We see from fig. 8 of ref. \cite{26} that $s_0 \simeq 1.20 - 1.25 \,
\, GeV$. We take $s_0 = 1.3 \,\, GeV$ in our calculations which leads to 
slightly better results. Finally we make use that  $\chi_{22}^{T=2} \simeq 1$ 
as stated in ref. \cite{26}.

For the phase shifts $\delta_{20}$ we make use of the fact that 
this channel is dominated by the $f_2 (1270)$ resonance (see {\bf section
4} for amplitudes of resonant terms). In fig. 6 we show the phase shifts
of this channel calculated from the resonant amplitude compared with 
experiment and the agreement with the data is good.

\begin{figure}[h]
\centerline{
\protect
\hbox{
\psfig{file=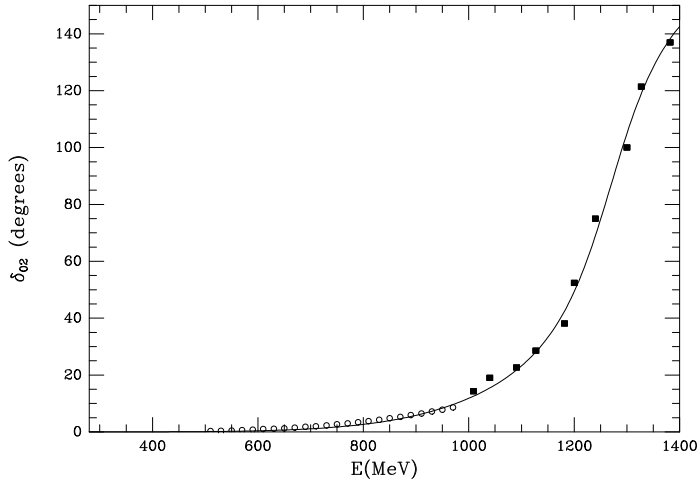,width=0.6\textwidth,angle=-90}}}
\caption{$\delta_{TL}=\delta_{02}$ phase shifts for the elastic $\pi\pi$ 
amplitude with $L=2$ and $T=0$ from threshold to $1.4$ $GeV$. This curve is 
given only through the exchange of the $f_2(1270)$ resonance. The 
experimental data are from ref. \cite{32}, squared points and \cite{A.D.Martin} 
for the empty circles.}
\end{figure}

For the $\gamma \gamma \rightarrow K^+ K^-$ reaction the Born 
term contribution in $L = 2$ 
is small compared to the one in S-wave up to about $\sqrt{s} = 1.4 \,
\, GeV$, due 
to the large $K$ mass. Moreover, this contribution is further reduced 
with a similar formula as eq. (13) with $s_0$ around the mass of the 
lowest resonance in each partial wave ($s_0 \simeq 1.3$ $GeV$ corresponding 
to the mass of the $f_2$ ($T=0$) and $a_2$ ($T=1$) resonances) \cite{26,28}. This allows 
us to neglect this term to a good approximation in the range 
$ \sqrt{s} < 1.4 $ $GeV$ where we are interested. 

Given the fact that the result of eq. (13) is generated by dispersion
relations using empirical input, the amplitude of eq. (13) should also
take into account the $L = 2$ contribution which is generated by our
vector and axial resonance exchange in the crossed
channels. Hence we consider explicitly only
the S-wave part of the resonance exchange as discussed in {\bf section 3.1}.

Apart from this background terms in $L = 2$, we have the contribution of
the $L = 2$ resonances which we discuss below.

\section{Direct coupling to the $f_2 (1270)$ and $a_2 (1320)$ resonances.}

Here we follow a standard procedure to include the resonances in the
same way as done in ref. \cite{5}.

As it is usually done \cite{5,10} we consider only the $\lambda = 2$ contribution
and hence we have the parametrization

\begin{equation}
t^{(2, 2)}_R = i 16 \pi \sqrt{\frac{20 \pi}{\beta (s)}} BW (s) (\Gamma_{
\gamma \gamma}^{(2)})^{1/2} \, Y_{22} (\cos \theta, \varphi)
\end{equation}

\noindent
where $\Gamma_{\gamma \gamma}^{(22)}$ is the resonance decay width
into two photons with opposite helicity. Furthermore

\begin{equation}
B W (s) = \frac{m_R \sqrt{\Gamma (\sqrt{s}) Br (M \bar{M})}}{
s - m_R^2 + i m_R \Gamma (\sqrt{s})}
\end{equation}

\begin{equation}
\Gamma (\sqrt{s}) = \Gamma (m_R) \, \Biggl[  \frac{p (\sqrt{s})}{p (m_R)}
 \Biggr]  ^5 \,
\frac{m_R}{\sqrt{s}} \, \frac{h (\sqrt{s})}{h (m_R)}
\end{equation}

\begin{equation}
h (\sqrt{s}) \, \propto [9 + 3 (p r)^2 + (p r)^4]^{-1}
\end{equation}

\noindent
where $h (\sqrt{s})$
is a decay form factor \cite{35}, $r$ is the effective interaction radius
taken as $r \simeq 1 \, fm$, $p$ is the CM momentum of the $M \bar{M}$ system
and
$\Gamma (\sqrt{s})$ the total width of the resonance, $Br (M \bar{M})$
is  the
branching ratio for the decay into the $M \bar{M}$ system such that
$\Gamma (\sqrt{s}) \, Br (M \bar{M}) S_i$ is the partial decay
width into the $M \bar{M}$ channel, with $S_i = 1/2, 1$
depending on whether the final $M \bar{M}$ state contains or does
not contain two identical particles. Given the fact that there is 
an important
interference for the $\gamma \gamma \rightarrow \pi^+ \pi^-$
reaction between the Born term in the $(2,2)$ channel and the
contribution of the resonance $f_2 (1270)$, it is then important that
the total (2,2) amplitude is properly unitarized and this is 
accomplished with the modification of the Born piece used in eq. (13). 
On the other hand, the choice of a constant value $r=1$ fm reproduces 
well the resonance around the peak but overestimates the tail of the 
resonance. We have chosen $r$ energy dependent in order to reproduce 
the T=0, L=2 phase shifts in terms of the $f_2(1270)$ resonance. We take 
$r=(\sqrt{s}-2 \; m_{\pi})/9 \; m_{\pi}$ which as can be 
seen in Fig. 6 reproduces very well the phase shifts $\delta _{02}$.

We take the following parameters for the resonances by means of which the
resonance strength, position and widths are well reproduced:

\vspace{0.5cm}

$a) f_2 (1270)$

\begin{equation}
\begin{array}{llll}
m_R = 1275 \, MeV &  \Gamma_R = 185 \, MeV &  Br (\pi\pi)
\Gamma_{\gamma \gamma}^{(2)} = 2.64 \,  KeV &
 Br (K \bar{K}) \Gamma_{\gamma \gamma}^{(2)} = 0.1 \, KeV
\end{array}
\end{equation}

\vspace{0.2cm}

$b) a_2 (1320)$

\begin{equation}
\begin{array}{llll}
m_R = 1318 \, MeV &  \Gamma_R = 105 \, MeV &  Br (\eta \pi)
\Gamma_{\gamma \gamma}^{(2)} = 0.19 \, KeV &
 Br (K \bar{K}) \Gamma_{\gamma \gamma}^{(2)} = 0.1 \, KeV
\end{array}
\end{equation}

These values are compatible with those of the Particle Data Group
\cite{36}.

\section{Final amplitudes for the $\gamma \gamma \rightarrow M \bar{M}$
reaction}

Let us introduce some notation in order to proceed to sum the different 
amplitudes. We denote by $\tilde{t}_{\chi \pi} (\tilde{t}_{\chi K})$ 
the chiral amplitude 
to order
$O (p^4)$ with charged $\pi (K)$ intermediate states from eq. (5) eliminating
the strong amplitude to order $O (p^2)$ which factorizes the amplitude

\begin{equation}
\tilde{t}_{\chi \pi} = \, - \frac{\epsilon_1 \epsilon_2}{16 \pi^2} \, 
2 e^2 \; 
\left\{ 1 + \frac{m_\pi^2}{s} \, [ l n \left( \frac{1 + (1 - 4 m_\pi^2 
/ s)^{\frac{1}{2}}}
{1 - (1 - 4 m_\pi^2/s)^{\frac{1}{2}}} \right
) - i \pi]^2 \right\}
\end{equation}

\noindent
with the obvious change $m_\pi^2 \rightarrow m_K^2$ for $\tilde{t}_{\chi K}$
and its analytical extrapolation below threshold.

We denote by $t_R (R \equiv \rho, \omega, A)$ the tree level resonance on
shell contributions in S-wave, which are given in refs. \cite{17,18}. In our
normalization these will be

\begin{equation}
\begin{array}{l}
t_{\rho} = 2 e^2 \epsilon_1 \epsilon_2 
R_{\rho} [ \frac{\displaystyle{m_\rho^2}}{\displaystyle{\beta (s)}} l n 
\left( \frac{\displaystyle{ 1 + \beta (s) + \frac{s_\rho}{s}}}
{\displaystyle{1 - \beta (s) + \frac{s_\rho}{s}}}
\right) - s] \\[4ex]
t_{\omega} = 2 e^2 \epsilon_1 \epsilon_2 
R_{\omega} [ \frac{\displaystyle{ m_\omega^2}}{\displaystyle{\beta (s)}} l n 
\left( \frac{ \displaystyle{ 1 + \beta (s) + \frac{s_\omega}{s}}}
{\displaystyle{ 1 - \beta (s) + \frac{s_\omega}{s}}}
\right) - s] \\[4ex]
t_A = - 2 e^2 \epsilon_1 \epsilon_2 
\frac{\displaystyle{ (L^r_9 + L^r_{10})}}{\displaystyle{f^2}}
[ \frac{\displaystyle{s_A}}{\displaystyle{2 \beta (s)}} l n 
\left( \frac{\displaystyle{1 + \beta (s) + \frac{s_A}{s}}}
{\displaystyle{1 - \beta (s) + \frac{s_A}{s}}}
\right) + s] 
\end{array}
\end{equation}

\noindent
where $s_{\omega} = 2 (m_\omega^2 - m_\pi^2), 
s_{\rho} = 2 (m_\rho^2 - m_\pi^2), 
s_{A} = 2 (m_A^2 - m_\pi^2) $ for $\gamma \gamma \rightarrow \pi^+ \pi^-$
and 
$s_{A} = 2 (m_A^2 - m_K^2)$ for $\gamma \gamma \rightarrow K^+ K^-$.

Since we have direct coupling to the resonances $f_2$ and $a_2$ with helicity
2, it is convenient to work in the helicity basis for the amplitudes. By
taking

\begin{equation}
\begin{array}{ll}
\epsilon_+ = - \frac{1}{\sqrt{2}} (\epsilon_x + i \epsilon_y) &
\epsilon_-  =  \frac{1}{\sqrt{2}} (\epsilon_x - i \epsilon_y)
\end{array}
\end{equation}

\noindent
The Born amplitude of eq. (1) is then separated into

\begin{equation}
\begin{array}{l}
t^{(0)} = t^{\lambda =  0}_B = t^{++}_B = 
i \frac{2 e^2 (1 - \beta^2)}{1 - \beta^2 cos^2 \theta} \\ [2ex]
t^{(2)} = t^{\lambda =  2}_B = t_B^{+ -}  = 
- i \frac{2 e^2 \beta^2 e^{2 i \phi} \sin^2 \theta}{1 - \beta^2 cos^2 \theta}
\end{array}
\end{equation}

\noindent
where $\beta \equiv \beta (s) = (1 - 4 m^2 / s)^{1/2}$ with $m$ the mass
of the corresponding charged meson.

The amplitudes which we have introduced $t_R,\tilde{t}_{\chi \pi},
\tilde{t}_{\chi K}, \tilde{t}_{R, M_1 M_2}$ only contribute
in S-wave and hence only have helicity zero. Thus, by means of eq. (23)
we have $t^{+ +} = i t$ where $t$ stands here for the different amplitudes.

Note that in $t^{+ +}$ one has contributions from $(J, \lambda) = (0, 0),
(2, 0)$... . However, we also mentioned that the $(2, 0)$ component of the
Born term is negligible. On the other hand following ref. \cite{26} this
Born component, plus corrections from crossed channels, is given by an
equation like eq. (13) and then the relative smallness remains. Hence we 
neglect this kind of contributions and similarly for the $t^{+-}$ amplitude 
for partial waves higher than 2.

For simplicity we take the full $t^{+-}_B$ amplitude in eq. (13)
rather than $t_B^{(2,2)}$ since the higher multipoles (4,2), (6,2), ... 
contributions are also very small. We will call $t_{BC}^{(2)}$ this amplitude.
The amplitudes used below are the physical amplitudes with proper 
normalization of the states, not the ``unitary'' normalization amplitudes
of eq. (8), (10) or those of ref. \cite{15} for symmetrical pionic states.
Note also that $t_A , t_B, t_f^{(2,2)}, t_a^{(2,2)}$ depend on the
channel as stated in eqs. (22), (24), (15-18). 

After this discussion we can already write the amplitudes:

\noindent

$1) \; \gamma \gamma \rightarrow \pi^+ \pi^- $

\begin{equation}
\begin{array}{l}
 t^{(0)} = t^{(0)}_B + i \{ t_A + t_\rho + (\tilde{t}_{A \pi^+ \pi^-}
+ \tilde{t}_{\rho \pi^+ \pi^-} + \tilde{t}_{\chi \pi}) t_{\pi^+ \pi^- ,
\pi^+ \pi^- }\, + \\
 + \, (\tilde{t}_{A K^+ K^-} + \tilde{t}_{\chi K}) t_{K^+ K^-,
\pi^+ \pi^-} + (\tilde{t}_{\rho \pi^0 \pi^0}+
\tilde{t}_{\omega \pi^0 \pi^0}) t_{\pi^0 \pi^0, \pi^+
\pi^-} \}\\

t^{(2)} = t^{(2)}_{BC} + t_{f_2}^{(22)}
\end{array}
\end{equation}

\noindent
where

\begin{equation}
\hspace{-6cm}
\begin{array}{l}
t_{\pi^+ \pi^- , \pi^+ \pi^-} = \frac{1}{3} t^{I = 0} + \frac{1}{6}
t^{I = 2}\\
t_{\pi^0 \pi^0 , \pi^+ \pi^-} = \frac{1}{3} t^{I = 0} - \frac{1}{3}
t^{I = 2}\\
t_{K^+  K^- , \pi^+ \pi^-} = \frac{1}{\sqrt{6}} t^{I = 0}
\end{array}
\end{equation}

\noindent
$2) \gamma \gamma \rightarrow \pi^0 \pi^0$

\begin{equation}
\hspace{2cm}
\begin{array}{l}
- i t^{(0)} = t_\omega + t_\rho+(\tilde{t}_{\rho \pi^0 \pi^0}
 +\tilde{t}_{\omega \pi^0 \pi^0} )
t_{\pi^0 \pi^0, \pi^0 \pi^0} + \\ + (\tilde{t}_{A \pi^+ \pi^-}
+ \tilde{t}_{\rho \pi^+ \pi^-} + \tilde{t}_{\chi \pi}) t_{\pi^+ \pi^-,
\pi^0 \pi^0} 
+(\tilde{t}_{\chi K} + \tilde{t}_{A K^+ K^-}) t_{K^+ K^-, \pi^0 \pi^0}\\

t^{(2)} = t_{f_2}^{(2,2)}
\end{array}
\end{equation}

\begin{equation}
\hspace{-6.5cm}
\begin{array}{l}
t_{\pi^0 \pi^0, \pi^0 \pi^0} = \frac{1}{3} t^{I = 0} + \frac{2}{3}
t^{I = 2}\\
t_{K^+ K^-, \pi^0 \pi^0} = \frac{1}{\sqrt{6}} t^{I = 0}\\
t_{\pi^+ \pi^-, \pi^0 \pi^0} = t_{\pi^0 \pi^0, \pi^+ \pi^-}
\end{array}
\end{equation}

\noindent
3) $\gamma \gamma \rightarrow \pi^0 \eta$

\begin{equation}
\hspace{-5cm}
\begin{array}{l}
- i t^{(0)} = ( \tilde{t}_{AK^+ K^-} + \tilde{t}_{\chi K}) t_{K^+ K^-,
\pi^0 \eta}\\

t^{(2)} = t^{(2,2)}_{a_2}
\end{array}
\end{equation}

\begin{equation}
\hspace{-6.3cm} t_{K^+ K^-, \pi^0 \eta} = -\frac{1}{\sqrt{2}} t^{I= 1}
\end{equation}

\noindent
4) $\gamma \gamma \rightarrow K^+ K^-$

\begin{equation}
\hspace{-0.3cm}
\begin{array}{l}
t^{(0)} = t_B^{(0)} + i \{ t_A + (\tilde{t}_{A \pi^+ \pi^-} +
\tilde{t}_{\rho \pi^+ \pi^-}  + \tilde{t}_{\chi \pi}) t_{\pi^+ \pi^- ,
K^+ K^-} + \\
+ (\tilde{t}_{\rho \pi^0 \pi^0}+\tilde{t}_{\omega \pi^0 \pi^0})
  t_{\pi^0 \pi^0, K^+ K^-} +
(\tilde{t}_{A K^+ K^-} + \tilde{t}_{\chi K}) t_{K^+ K^- , K^+ K^-} \} \\

t^{(2)} =  t^{(2,2)}_{f_2} + t^{(2,2)} _{a_2}
\end{array}
\end{equation}

\begin{equation}
\hspace{-5.9cm}
\begin{array}{l}
t_{\pi^+ \pi^-, K^+ K^-} = t_{K^+ K^-, \pi^+ \pi^-}\\
t_{\pi^0 \pi^0, K^+ K^-} = t_{K^+ K^-, \pi^0 \pi^0}\\
t_{K^+ K^-, K^+ K^-} =  \frac{1}{2} t^{I = 0} + \frac{1}{2}
t^{I = 1}
\end{array}
\end{equation}

\noindent
$5) \gamma \gamma \rightarrow K^0 \bar{K}^0$

\begin{equation}
\hspace{-1.cm}
\begin{array}{l}
- i t^{(0)} = (\tilde{t}_{A \pi^+ \pi^-} + \tilde{t}_{\rho \pi^+ \pi^-} +
\tilde{t}_{\chi K}) t_{\pi^+ \pi^-, K^0 \bar{K}^0} +\\
+( \tilde{t}_{\rho \pi^0 \pi^0}+ \tilde{t}_{\omega \pi^0 \pi^0})
 t_{\pi^0 \pi^0, K^0 \bar{K}^0} +
(\tilde{t}_{A K^+ K^-} + \tilde{t}_{\chi K}) t_{K^+ K^-, K^0 \bar{K}^0}\\

t^{(2)} = t^{(2,2)}_{f_2} - t^{(2,2)} _{a_2}
\end{array}
\end{equation}

\begin{equation}
\hspace{-6.2cm}
\begin{array}{l}
t_{\pi^+ \pi^-, K^0 \bar{K}^0} = t_{K^+ K^-, \pi^+ \pi^-}\\
t_{\pi^0 \pi^0, K^0 \bar{K}^0} = t_{K^+ K^-, \pi^0 \pi^0}\\
t_{K^+ K^-, K^0 \bar{K}^0} =  \frac{1}{2} t^{I = 0} - \frac{1}{2}
t^{I = 1}
\end{array}
\end{equation}

It is interesting to note that the isoscalar $f_2$ and isovector $a_2$ 
resonances interfere constructively in $\gamma \gamma \rightarrow 
K^+ K^-$ and destructively in $\gamma \gamma \rightarrow K^0 \bar{
K}^0 $ as shown at the end of eqs. (31) and (33). This fact has been predicted 
\cite{lip} long before its observation (see experimental results 
in figs. 11 and 12). 

The width for $\omega \rightarrow \gamma \eta $ is about two orders of magnitude 
smaller than for $\omega \rightarrow \gamma \pi^0$, resulting in a smaller 
coupling. For this reason we omit the contributions involving the $\omega 
\rightarrow \gamma \eta$ coupling which should appear in $\gamma \gamma 
\rightarrow \pi^0 \eta$, $K \bar{K}$. We have, however, estimated the 
relevance of this term calculating the tree level and the final state 
interaction correction (only the imaginary part of this latter term) for the 
$\gamma \gamma \rightarrow \pi^0 \eta$ channel, where it is more relevant.
 We have seen that it gives small corrections which will be shown in 
 {\bf section $6$}.

\section{Differential and integrated cross sections for 
$\gamma \gamma \rightarrow M \bar{M}$}

In terms of the $t^{(0)}$ and $t^{(2)}$ amplitudes we have

\begin{equation}
\frac{d \sigma}{d \cos \theta} = \frac{\beta}{64 \pi s}
[| t^{(0)}|^2 + |t^{(2)}|^2]
\end{equation}

\noindent
and the cross section integrated for $| \cos \theta| < Z$, as given
in some experiments,

\begin{equation}
\sigma = 2 \int_0^Z \frac{d \sigma}{d \cos \theta} \, d \cos \theta
\end{equation}

\begin{figure}[h]
\centerline{
\protect
\hbox{
\psfig{file=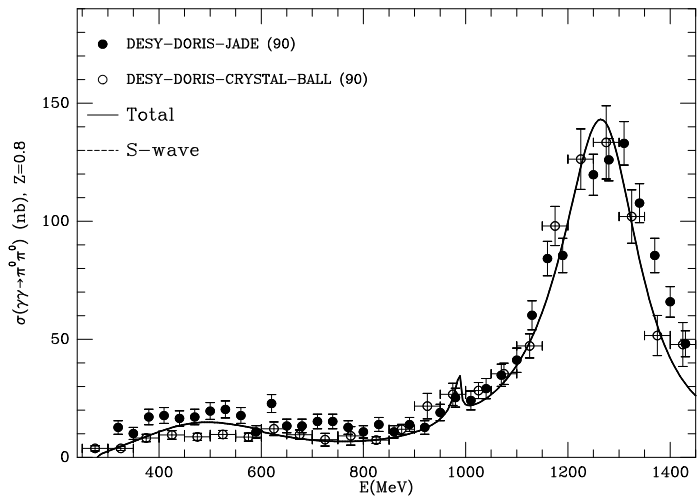,width=0.6\textwidth,angle=-90}}}
\caption{The integrated cross section for $\gamma\gamma\rightarrow \pi^0\pi^0$ 
with $Z=0.8$ compared with the data from ref. \cite{1} and ref. \cite{
oest}, the latter normalized in the $f_2(1270)$ peak region.}
\end{figure}

In fig. 7 we show the cross section for $\gamma \gamma \rightarrow
\pi^0 \pi^0$ integrated up to $Z = 0.8$ compared to the data
of Crystal Ball  \cite{1} and JADE \cite{oest}. The data of Crystal 
Ball are normalized while those of JADE correspond to an unnormalized 
distribution. We have chosen a normalization of this latter data such that in 
the large peak of the cross section corresponding to the $f_2(1270)$ 
resonance the two cross sections have the same strength. Our results 
are in agreement with those of the 
Crystall Ball experiment at very low energies, where they agree with 
those of the two loop calculations in $\chi PT$ \cite{22}, and for 
$\sqrt{s} > 0.6$ $GeV$. For $0.4 < \sqrt{s} < 0.6$ our results lie 
between those of the Crystall Ball and JADE experiments. The calculation 
shows a broad bump in this region, a consequence of the presence of the 
$\sigma$ meson pole in the L=0, T=0 channel around $(470+i200)$ $MeV$. The 
data around the $f_2$ resonance is well reproduced.

It is quite interesting to see that the $f_0$ resonance shows up 
as a small peak in the cross section, in the lines of what is observed 
in both experiments. The smallness of the peak in our calculation 
is due to interference with the $\omega$ contribution.

\begin{figure}[H]
\centerline{
\protect
\hbox{
\psfig{file=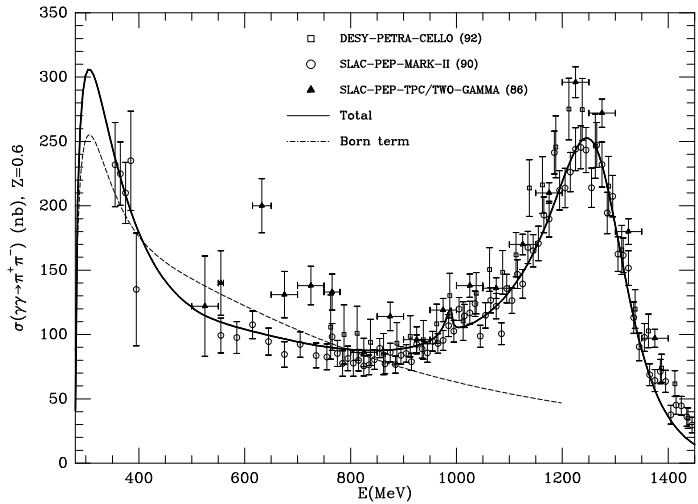,width=0.6\textwidth,angle=-90}}}
\caption{The integrated cross sections for  $\gamma\gamma\rightarrow \pi^+\pi^-$ 
with Z=0.6 compared with the experimental data from refs. \cite{4,3,2}
 respectively 
from up to down in the figure as indicated in the text. We also show the Born 
term calculation.}
\end{figure}

In fig. 8 we show the cross section for $\gamma \gamma \rightarrow \pi^+
\pi^-$ integrated up to $Z = 0.6$ compared to the experimental data of the
SLAC-PEP-TPC/ TWO GAMMA \cite{2}, SLAC-PEP-MARK II \cite{3} and 
DESY-PETRA-CELLO \cite{4} collaborations. The agreement with the data is 
good in general particularly with the results of MARK-II. We can see that 
the $f_0$ resonance does not show up significantly in the data nor 
in the calculation. Note also that we reproduce the peak corresponding 
to the $f_2(1270)$ both in figs. 7 and 8.

\begin{figure}[H]
\centerline{
\protect
\hbox{
\psfig{file=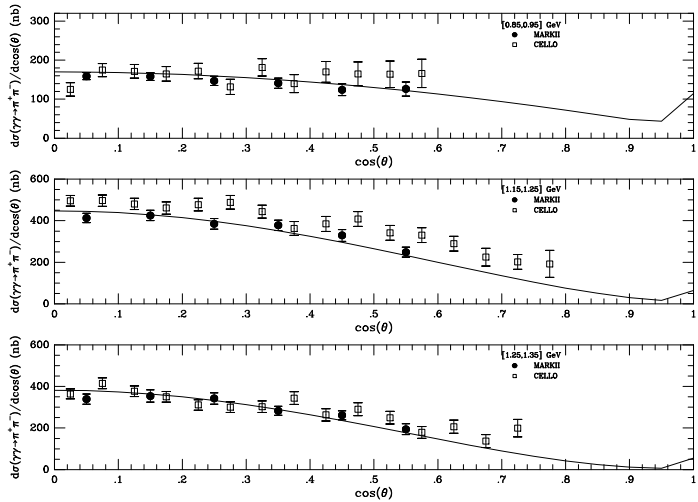,width=0.6\textwidth,angle=-90}}}
\caption{ Differential cross sections for  $\gamma\gamma \rightarrow
 \pi^+\pi^-$ in three energy regions compared with the data of MARKII
\cite{3} and CELLO\cite{4} around the $f_2(1270)$ resonance peak.}
\end{figure}

In figs. 9a,b,c we present differential cross sections for 
$\gamma \gamma \rightarrow \pi^+ \pi^-$ in three different energy regions 
around the peak of the $f_2(1270)$ resonance.
In general one observes a good agreement with the data. Note also that 
these differential cross sections were used in ref. \cite{27} together 
with the former cross section in order to
justify the existence of the resonance $f_0 (1100)$. We see that we 
reproduce the data without the need to introduce this resonance. The use
of our precise L = 0 amplitudes and the unitary scheme followed here
produce the necessary S-wave contribution to weaken the angular dependence
of the cross sections as observed by experiment.

\begin{figure}[H]
\centerline{
\protect
\hbox{
\psfig{file=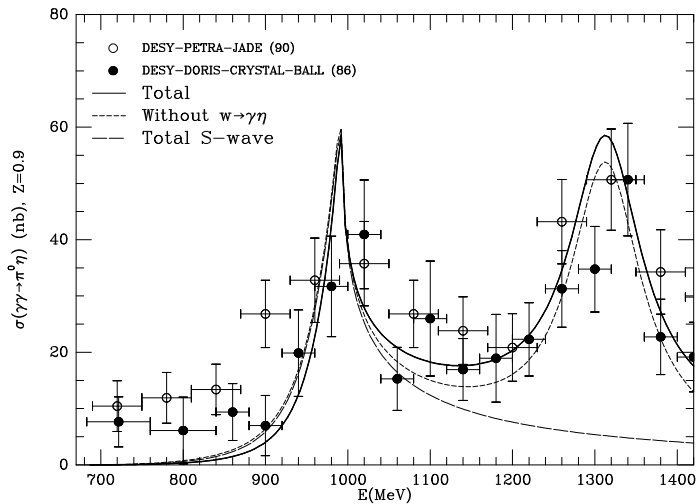,width=0.6\textwidth,angle=-90}}}
\caption{Integrated cross section for $\gamma \gamma \rightarrow \pi^0\eta$ 
with $Z=0.9$ $GeV$ compared with the experimental data from ref. \cite{oest} 
and ref. \cite{8}, the latter normalized in the $a_2(1320)$ peak region. 
We show our calculations for the S-wave contribution. The results with 
and without the $\omega$ contribution are also shown.}
\end{figure}

In fig. 10 we show our results for the $\gamma\gamma \rightarrow \pi^0 
\eta$ compared to the data of DESY-PETRA-JADE \cite{oest} and 
DESY-DORIS-CRYSTALL-BALL \cite{8} ($Z=0.9$). The data of ref. \cite{8} are 
normalized but those of ref. \cite{oest} have arbitrary normalization. 
In the figure we have normalized them such that they have the same 
strength as those of ref. \cite{8} in the $a_2(1320)$ peak. We can see a 
fair agreement of our results with the data, both around the $a_2(1320)$ 
resonance (the parameters of which are taken from the particle data 
tables) and for the peak around the $a_0(980)$ resonance, which 
results naturally from the use of our unitarized meson-meson 
amplitudes. We also show in the figure the S-wave 
contribution alone, which includes the $a_0(980)$ peak. We 
reproduce the data in the intermediate energy between the two 
resonances without the need to introduce an additional background, which 
has been sometimes assumed to come from a broad $a_0(1100-1300)$ 
resonance \cite{10}. In the figure we also show the result including 
the contribution of the exchange of the $\omega$, estimated as indicated 
in {\bf section 2}. We see negligible corrections in the peak 
of the $a_0$ and more significative changes around the minimum of the 
cross section.

In \cite{29} the peak of the $a_0$ and the deep region were also 
reproduced without the inclusion of a background. However, as noted 
in ref. \cite{10} the $\gamma \gamma \rightarrow K^+K^-$ amplitude 
was overestimated since the Born amplitude was used which is 
drastically reduced due to final state interaction, as we shall 
see below. On the other hand the exchange of an axial resonance, 
as we do here, was not included in ref. \cite{29} and this is 
essential to reproduce the strength of the $a_0(980)$ peak. Indeed 
if we take $L_9^r+L_{10}^r=0$ we get a strength of the peak around 
one third of the experimental strength.

\begin{figure}[H]
\centerline{
\protect
\hbox{
\psfig{file=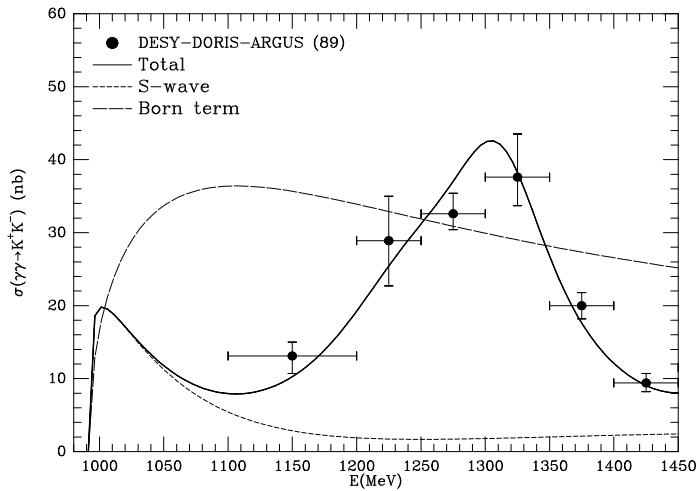,width=0.6\textwidth,angle=-90}}}
\caption{Integrated  $\gamma\gamma\rightarrow K^+ K^-$ from threshold to 1.45 
$GeV$ compared with the experimental data from ref. \cite{5}. We also show 
the Born contribution and the S-wave background.}
\end{figure}

In fig. 11 we show results for $\gamma \gamma \rightarrow K^+ K^-$ together
with the data of DESY-DORIS-ARGUS \cite{5}. We also show there the 
contribution of the Born term without corrections and the background in
S-wave. The cross section is also well reproduced in this channel. The most
striking feature in this figure is the drastic reduction of the Born term
contribution due to FSI and to the crossed channel contributions. 
Around $\sqrt{s} = 1070 \; \, MeV$
 this reduction is more than a factor 
ten. The need to reduce drastically the Born contribution had been pointed
out
before but no theoretical justification had been found so far \cite{10}.

\begin{figure}[H]
\centerline{
\protect
\hbox{
\psfig{file=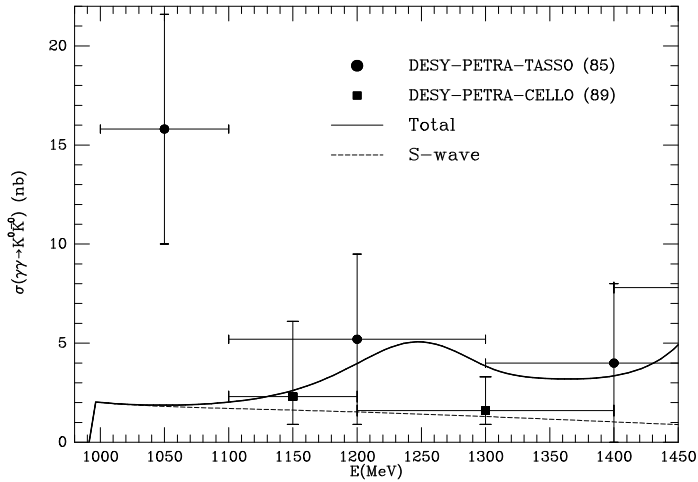,width=0.6\textwidth,angle=-90}}}
\caption{Integrated  $\gamma\gamma\rightarrow K^0 \bar{K}^0$ from threshold to 
1.45 $GeV$ compared with the data from ref. \cite{6,7} respectively as it is 
indicated in the figure from up to down. We also show the S-wave contribution.}
\end{figure}

In fig. 12 we show the results for $\gamma \gamma \rightarrow K^0 \bar{K}^0$
compared to the data of DESY-PETRA-TASSO \cite{6} and DESY-PETRA-CELLO
\cite{7}. The results are compatible with the data,
which, however, have large errors.
We also show the background of S-wave without the contribution
of the $f_2$ and $a_2$ resonances. 
The background found is small as expected, but not because of the lack of a
Born term, but because of cancellations between important contributions which
were also responsible for the reduction of the 
$\gamma \gamma \rightarrow K^+ K^-$ Born term.

\section{Partial decay width to two photons of the $f_0 (980)$
and $a_0 (980)$.}

In ref. \cite{15} the partial decay widths of the $f_0 (980)$ and 
$a_0 (980)$ resonances into $\pi \pi, K \bar{K}$ or $\pi^0 \eta$ were 
evaluated. In this section we complete the information evaluating the partial
decay widths into $\gamma \gamma$.

From our amplitudes in eqs. (30) in isospin $T = 1$ and eq. (32) for the
isospin $T = 0$ part, by taking the terms which involve the strong
$M \bar{M} \rightarrow M \bar{M}$ amplitude, we isolate the part of the
$\gamma \gamma \rightarrow M \bar{M}$ which proceeds via the resonances $a_0$
and $f_0$ respectively. In the vicinity of the resonance the amplitude 
proceeds
as $M \bar{M} \rightarrow R \rightarrow M \bar{M}$. Then we eliminate
the $R \rightarrow M \bar{M}$ part of the amplitude plus the $R$ 
propagator and remove the proper
isospin Clebsch Gordan coefficients for the final states (1 for $\pi^0 \eta$
and $- 1/ \sqrt{2}$ for $K^+ K^-$) and then we get the coupling of the 
resonances to the $\gamma \gamma$ channel. It is convenient to do this for
the $K^+ K^-$ final states because in the case of the 
pions one has a large background of the $\sigma$ resonance in the elastic
amplitude.

\begin{figure}[h]
\centerline{
\protect
\hbox{
\psfig{file=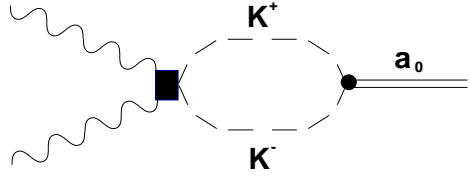,width=0.4\textwidth,angle=-90}}}
\caption{ $g_{a\gamma\gamma}$ through $K^+ K^-$ intermediate states, connecting 
the $a_0$ resonance with the two photons.}
\end{figure}

Diagrammatically this is represented in fig. 13 for the $a_0$ case. The 
$g_{a_0 \gamma \gamma}$ coupling is then given by

\begin{equation}
g_{a_0 \gamma \gamma} = - \frac{g_{a_0 K \bar{K}}}{\sqrt{2}} \,
[\tilde{t}_{\chi K} + \tilde{t}_{A K^+ K^-}]
\end{equation}

\noindent
where $g_{a_0 \gamma \gamma}$ is the coupling of the $a_0$ to the $K \bar{K}$
system evaluated in ref. \cite{15}. The $- 1 / \sqrt{2}$ in front of eq.
(38) is the Clebsch Gordan coefficient of $|K \bar{K}, T = 1>$ to
$K^+ K^-$. Following ref. \cite{15} the decay width of $a_0 \rightarrow
\gamma \gamma$ is given by

\begin{equation}
\Gamma_{a_0} ^{\gamma \gamma} = \frac{1}{16 \pi^2} \, \frac{1}{2} \,
\sum_{\lambda_1, \lambda_2} \, \int^{\infty}_0 \, d W \, \frac{q}{W^2} \,
|g_{a_0 \gamma \gamma}|^2 \, 
\frac{\Gamma_{a_0} (W)}{(M_{a_0} - W)^2 + (\frac{\Gamma_{a_0}
 (W)}{2})^2}
\end{equation}

\noindent
where $\lambda_1 , \lambda_2$ are the photon polarizations and $W = \sqrt{s}$.
The coupling $g_{a_0 K \bar{K}}$ is evaluated in ref. \cite{15} in terms
of $Im \, t_{11}$ of $K \bar{K} \rightarrow K \bar{K}, T = 1$ amplitude and
hence we find

\begin{equation}
\Gamma_{a_0} ^{ \gamma \gamma } = \frac{- 1}{16 \pi^2} \, 
\int^{\infty}_0 \, d W \, \frac{q}{W^2} \,
4 M_R \,
|\frac{\tilde{t}_{\chi K} + \tilde{t}_{A K^+ K^-}}{\sqrt{2}}|^2 \,
Im \, t_{11}
\end{equation}

\noindent
where the lower limit in practice is the threshold for the lightest 
$M \bar{M}$ pair in this channel where $I m t_{11} \neq 0$ ($\pi^0 \eta$
in this case). The integral extends over $W$ about
$2 \, \Gamma$ above and below $M_R$, the pole mass $M_{a_0} = 1009 \;
\, MeV$, with $\Gamma=112$ $MeV$. We thus obtain

\noindent

\begin{equation}
\Gamma_{a_0}^{\gamma \gamma} = 0.78 \; KeV
\end{equation}

\noindent
and the related quantity

\begin{equation}
\Gamma_{a_0}^{\gamma \gamma} 
\frac{\Gamma_{a_0}^{\eta \pi}}{\Gamma_{a_0}^{tot}}
= 0.49 \; \, KeV
\end{equation}

For the $f_0$, apart from $K^+ K^-$ intermediate states we have $\pi^+ \pi^-$,
and also $\pi^0 \pi^0$ through the $\omega$ exchange. Thus we have

\begin{equation}
g_{f_0 \gamma \gamma} = - \frac{g_{f_0 K \bar{K}}}{\sqrt{2}} 
(\tilde{t}_{A K^+ K^-} + \tilde{t}_{\chi K})
- \frac{g_{f_0 \pi \pi}}{\sqrt{3}} 
(\tilde{t}_{A \pi^+ \pi^-} + \tilde{t}_{\rho \pi^+ \pi^-} + 
\tilde{t}_{\chi \pi} + \tilde{t}_{\rho \pi^0 \pi ^0} + 
\tilde{t}_{\omega \pi^0 \pi ^0})
\end{equation}

Hence by writing the strong $g_{f_0 K \bar{K}}$ and $g_{f_0 \pi\pi}$ couplings 
in terms of the strong amplitudes we find

\begin{equation}
\begin{array}{l}
\Gamma_{f_0}^{\gamma \gamma} = - \frac{1}{16 \pi^2} \,
\int^{\infty}_0 \, d W \, \frac{q}{W^2} \,
4 M_R \, \left\{ \right.
| \frac{\chi_K}{\sqrt{2}}|^2 I m t_{11} \\[2ex]
 + |\frac{\chi_{\pi}}{\sqrt{3}}|^2 \,
\frac{(Re(t_{21}))^2}{I m t_{11}} \, - \sqrt{\frac{2}{3}} \,
Re(t_{21}) \hbox{Im(} 
\chi_K \chi_{\pi}^*)
\left. \right\}
\end{array}
\end{equation}

\noindent
where for simplicity we have introduced the notation

\begin{equation}
\begin{array}{l}
\chi_K = \tilde{t}_{\chi K} + \tilde{t}_{A K^+ K^-} \\[2ex]
\chi_{\pi} = 
\tilde{t}_{A \pi^+ \pi^-} + \tilde{t}_{\rho \pi^+ \pi^-} + 
\tilde{t}_{\chi \pi} + \tilde{t}_{\omega \pi^0 \pi^0}+ \tilde{t}_{\rho 
\pi^0 \pi^0}
\end{array}
\end{equation}

In this case we also integrate in $W$ $2  \, \Gamma$ up and down $M_{f_0}
 (M_{f_0} = 992 \;\, MeV)$, with  $\Gamma_{f_0} = 25 \, \, MeV$ \cite{15}.

Thus we get

\begin{equation}
\Gamma^{\gamma \gamma}_{f_0} = 0.20 \;  \, KeV
\end{equation}

The result of eq. (41) is larger than the results quoted in the 
particle data group \cite{36}, ($0.28 \pm 0.04 \pm 0.1$) KeV from 
ref. \cite{oest} and ($0.19 \pm 0.07 \pm 0.10 $) KeV form ref. \cite{8}. 
However, one should note that in the experimental analysis a 
background term is assumed while in our approach the strength around 
the $a_0$ peak 
in the $\gamma \gamma \rightarrow \pi^0 \eta$ cross section comes 
from the $a_0$ excitation.

The result of eq. (45) is smaller than the average in the PDG 
\cite{36} ($0.56 \pm 0.11$) $KeV$, although consistent with some analyses, 
$(0.29 \pm 0.07 \pm 0.12)$ $KeV$ \cite{3} and $(0.31 \pm 0.14 \pm 0.09)$ 
$KeV$ \cite{1}.

Eq. (43) has made use of a peculiar property of the $f_0$ resonance which is 
that the $t_{21}$ amplitude around the $f_0$ resonance can be approximately 
described by an ordinary Breit-Wigner form multiplied by the phase factor 
$e^{i \pi/2}$. This can be seen from the experimental phase shifts for 
$K \bar{K} \rightarrow \pi\pi$, T=0, L=0, which lie around $220$ degrees, (
see Fig. 4 of \cite{15}). This fact was overlooked in the partial decay 
analysis of the $f_0$ resonance in \cite{15}. This deficiency, together 
with some small numerical corrections which lead to a slightly smaller 
cut off of $\Lambda=1$ $GeV$, have been taken into account in a reanalysis 
of these partial decay widths in \cite{BNL}. We use here this updated 
information.  

\section{Conclusions}

We have presented here a unified theoretical approach for the reaction
$\gamma \gamma \rightarrow M \bar{M}$ with 
$M \bar{M} = \pi^+ \pi^- , \pi^0 \pi^0 , K^0 \bar{K}^0 , \pi^0 \eta$. 
An important new ingredient with respect to other works is the treatment of
the S-wave $M \bar{M}$ amplitudes in $T = 0$, $1$, which we have taken 
from a recent successful work based upon chiral symmetry. This allows us to
treat accurately the strong final state interaction which plays a major role
in this reaction.

We have also taken into account well established facts concerning the role
of the exchange of vector and axial resonances in the $t$ and $u$ channels.

The direct coupling to the $f_2 (1270)$
and $a_2 (1320)$ resonances has been introduced explicitly in a standard way 
assuming dominance of the helicity two amplitudes, as customarily done.

With all these ingredients we study for the first time all these final 
mesonic states with a unified approach and obtain a general agreement in all
channels up to about $\sqrt{s} = 1.4 \, MeV$.

Some results of our study are worth stressing:

1) The resonance $f_0 (980)$ shows up weakly in 
$\gamma \gamma \rightarrow \pi^0 \pi^0$ and barely in 
$\gamma \gamma \rightarrow \pi^+ \pi^-$.

2) In order to explain the angular distributions of the
$\gamma \gamma \rightarrow \pi^+ \pi^-$  reaction we did not need the
hypothetical $f_0 (1100)$ broad resonance suggested in other works \cite{27}.
 This also solves the puzzle of why it did not show up in the 
$\gamma \gamma \rightarrow \pi^0 \pi^0$ channel. Furthermore, such resonance
does not appear in the theoretical work of ref. \cite{15}, while the 
$f_0 (980)$ showed up clearly as a pole of the $t$ matrix in $T = 0$.

3) The resonance $a_0$ shows up clearly in the 
$\gamma \gamma \rightarrow \pi^0 \eta$ channel and we reproduce 
the experimental results without the need of an extra background 
from a hypothetical $a_0(1100-1300)$ resonance suggested in ref. 
\cite{10}.

4) We have found an explanation to the needed reduction of the Born term
in the $\gamma \gamma \rightarrow  K^+ K^-$  reaction in terms of
final state interaction of the $K^+ K^-$ system.

5) In the case of $\gamma \gamma \rightarrow K^0 \bar{K}^0$ we find a small
cross section, which is not due to the lack of Born terms, but to a 
cancellation of magnitudes of the order of the
$\gamma \gamma \rightarrow K^+ K^-$ Born term.

\vspace{1cm}

{\bf Acknowledgements}

We would like to acknowledge fruitful discussions with  A. Pich, J. Prades, 
A. Bramon and F. Guerrero. One of us J. A. O. would like to 
acknowledge finantial support from the Generalitat Valenciana. This work 
is partially supported by CICYT, contract
no. AEN 96-1719.

\end{document}